# Sub-Diffraction Chromatin Domains: Architecture, Regulation, and Functional Roles in Nuclear Organization

Vinayak Vinayak[1,2], Melike Lakadamyali[1,3], Vivek B Shenoy[1,2]

[1] Center for Engineering Mechanobiology, University of Pennsylvania, USA
[2] Department of Materials Science and Engineering, University of Pennsylvania, USA
[3] Department of Physiology, Perelman School of Medicine, University of Pennsylvania, Philadelphia, USA

## Abstract

Nanoscale chromatin domains—variously termed nucleosome clutches, nanodomains, or packing domains—have emerged as fundamental architectural units of the mammalian genome during interphase and mitosis. Unlike cohesin-dependent loops or TADs, these 50–200 nm structures persist in the absence of loop extrusion, pointing to a distinct organizing principle shaped by histone post-translational modifications and constrained by interactions with the nuclear lamina. Super-resolution microscopy and electron tomography now enable their direct visualization, revealing conserved features such as fractal packing, enrichment for linker histone H1, and radial stratification of active and repressive histone marks. Accumulating evidence indicates that these domains act as transcriptional hubs, dynamically remodel in response to developmental and environmental cues, and undergo pathological disruption in disease. Integrated experimental, theoretical, and computational insights suggest that chromatin-protein interactions, epigenetic read–write processes, and diffusion-driven dynamics together govern their formation, persistence, and nuclear positioning. Viewed in this light, nanoscale domains represent a privileged regulatory tier, complementary to compartments and loop-based structures, that bridges local chromatin states with global nuclear architecture. By situating them alongside lamin-associated (LADs) and nucleolus-associated domains (NADs), we propose a unified biophysical framework for chromatin organization across scales and outline key open questions for future exploration. Because their structural disruption is a recurring feature of aging, cancer, and degenerative diseases, understanding these domains may open new avenues for diagnostics and therapeutic intervention.

## 1. Introduction

The spatial organization of chromatin within the nucleus is fundamental to genome regulation yet remains only partially understood. At the broadest scale, chromatin segregates into euchromatin and heterochromatin: two chromatin states distinguished by transcriptional activity, accessibility, and histone modifications. Euchromatin, characterized by high DNA accessibility and active marks such as H3K9ac and H3K27ac, supports gene expression, while compact, inaccessible, transcriptionally silent heterochromatin is enriched in repressive marks like H3K9me3 and H3K27me3 along with associations to proteins such as HP1[7, 20-23]. This binary framework has historically underpinned models of genome function, yet recent imaging and sequencing studies reveal additional layers of chromatin organization beyond this simple dichotomy.

Over the past decade, genome-wide chromosome conformation capture techniques (such as Hi-C[24] and Micro-C[25]), together with advances in fluorescence-based microscopy, have revealed hierarchical layers of three-dimensional genome folding. These include chromosome territories, A/B compartments, topologically associating domains (TADs), and chromatin loops, where the majority of loops are formed and stabilized by architectural proteins such as CTCF and cohesin. Concurrently, physics-based models have successfully explained the occurrence of these scales of organization through phase separation and the action of active motors[27, 28]. Altogether, these features, often derived from ensemble measurements, have transformed our understanding of how multiscale genome topology interacts with and affects transcriptional control and cell phenotype.

However, these refined models remain incomplete, constrained by diffraction-limited microscopy and the still-developing reliability of single-cell sequencing approaches. More recently, emerging single-molecule localization microscopy (SMLM), also known as, super-resolution imaging approaches have uncovered a striking layer of sub-diffraction chromatin organization: dense, spatially discrete, heterogenous domains ranging from ~50 to 200 nanometers[1, 29]. These nanoscale structures, consistently observed across mammalian cell types, display conserved biophysical and epigenetic characteristics that lend them distinct architectural features. Notably, they persist even in the absence of classical loop extrusion machinery highlighting a distinct mechanistic foundation. Their dynamic reorganization during development and disease further underscores their potential role in cell fate decisions. Yet, existing physics-based models only partially explain their emergence, suggesting that these domains may represent a privileged layer of genome regulation that complements established higher-order structures.

In this Review, we examine the growing body of evidence surrounding sub-diffraction chromatin domains. We begin with the imaging methodologies that enabled their discovery, then assess their physical and epigenetic architecture, aiming to establish a consistent, biophysically grounded framework. Next, we explore their interactions with the transcriptional machinery and their dynamic remodeling in response to environmental and developmental cues, including implications in disease and therapy through epigenetic drugs. Finally, we consider how these structures integrate with existing hierarchy of genome organization and highlight key conceptual and technical questions that remain. Through this synthesis, we aim to provide a cohesive framework for understanding the role of nanoscale chromatin architecture in genome function.

## 2. Super-resolution Imaging to visualize Sub-diffraction Domains

Building on recent advances in genome-wide conformation capture[30] and the conceptual framework of euchromatin-heterochromatin segregation, super-resolution imaging has uncovered a deeper stratum of chromatin architecture composed of nanoscale, sub-diffraction-limited domains ranging from ~50 to 200 nm[1]. Historically, chromatin was widely thought to fold into a canonical 30-nm fiber; however, this model has been largely overturned by super-resolution and electron microscopy studies, which revealed irregularly packed, flexible chains of nucleosomes with no consistent higher-order fiber[1, 5, 31, 32]. This conceptual shift opened the door to investigating chromatin organization beyond diffraction-limited resolution. A convergence of imaging techniques, including super-resolution fluorescence microscopy, correlative electron imaging, and multiplexed genomic FISH, has revealed that these domains are conserved across mammalian cell types (Table 1 and Fig. 1). Their robust spatial characteristics and independence from topological constraints of looping machinery suggest they represent a mechanistically distinct level of genome architecture that complements established models of multiscale nuclear organization.

> **Box 1: How Single-Molecule Localization microscopy (SMLM) works: a physics perspective**
>
> **The Diffraction Barrier**: In a conventional light microscope, two molecules closer than ~200–300 nm blur into one spot because of the *diffraction limit*. This limit comes from the wave nature of visible light: when light passes through a circular lens aperture it diffracts, producing an Airy disk pattern rather than a perfect point. The smallest resolvable separation is given approximately by Abbe's formula, $d \approx \lambda/(2 \times NA)$, where $\lambda$ is the wavelength of visible light (~400–700 nm) and NA is the numerical aperture of the lens. As a result, overlapping wavefronts from objects closer than this distance cannot be distinguished as separate points. *An everyday analogy is two distant car headlights: if they are too close together, diffraction makes them blur into a single bright spot.*
>
> **Principle of SMLM[10]**: The insight behind SMLM (which encompasses STORM[18], PALM[19], DNA-PAINT[26] and other similar techniques) is that it is not necessary to resolve all molecules at once; only their individual positions need to be determined. Instead of all fluorophores emitting simultaneously, only a sparse random subset is switched "on" at a time. Each active molecule appears as a blurred spot (a point-spread function), which can be fitted mathematically to determine its center with nanometer accuracy. This localization accuracy depends on photon statistics and improves with the number of detected photons $N$ (error scales as $1/\sqrt{N}$). *This is like listening to people in a noisy room: if everyone speaks at once you cannot distinguish voices, but if only one person speaks at a time you can pinpoint their location with great accuracy and the greater number of times they speak, the more certain you are about their position.* By repeating many cycles of stochastic activation and localization, a high-resolution image is reconstructed from the accumulated coordinates. Typical SMLM experiments achieve ~20 nm resolution, an order of magnitude beyond conventional light microscopy.
>
> **Recognition and impact**: The transformative significance of super-resolution methods was acknowledged by the 2014 Nobel Prize in Chemistry, awarded to Eric Betzig, Stefan Hell, and William Moerner and followed by the 2019 Breakthrough Prize in Life Sciences to Xiaowei Zhuang. In chromatin biology, STORM directly revealed that nucleosomes organize into heterogeneous "clutches" and nanometer-scale domains, overturning the classical 30-nm fiber view and establishing the nanoscale architecture that underpins genome regulation.

Super-resolution Fluorescence Imaging Reveals Heterogeneous Nanodomains: Initial insight came from Stochastic Optical Reconstruction Microscopy (STORM) imaging of histone H2B in mammalian nuclei by Ricci et al., who showed that nucleosomes organize into heterogeneous, nanoscale clusters or "clutches" varying in size and density depending on cell pluripotency state[1]. Larger and denser clutches were associated with heterochromatin and excluded RNA polymerase II (RNA polII) suggesting that they have lower transcriptional activity, while smaller, more dispersed clutches co-localized with RNA polII, suggesting active transcription sites. Simultaneously, the discovery of the domains was confirmed by Prakash et al., who used super-resolution localization microscopy on meiotic pachytene chromosomes to demonstrate periodic arrangements of active and repressive histone marks, further underscoring nanoscale epigenetic compartmentalization[29].

In live cells, Nozaki et al. applied photoactivated localization microscopy (PALM) to visualize chromatin domains that appeared as compact, dynamic structures ~160–220 nm in diameter. These domains moved coherently and persisted through mitosis, acting as stable building blocks of chromosome organization[3]. Notably, their structural organization was influenced by cohesin and nucleosome-nucleosome interactions.

Xu et al. used STORM imaging to classify chromatin domains based on histone modifications, revealing three distinct spatial patterns: segregated nanoclusters (histone acetylation), dispersed nanodomains (active methylation), and compact aggregates (repressive methylation)[2]. Active transcription was spatially correlated with less compact chromatin and repressive marks were found in densely packed regions[2, 7]. Rahman et al. expanded the scale and dimensionality of super-resolution nuclear profiling by implementing 3D multiplexed Exchange-PAINT and HIST microscopy, showing that proteins and chromatin marks organize into nanoscale domains with distinct spatial scaling behavior[15].

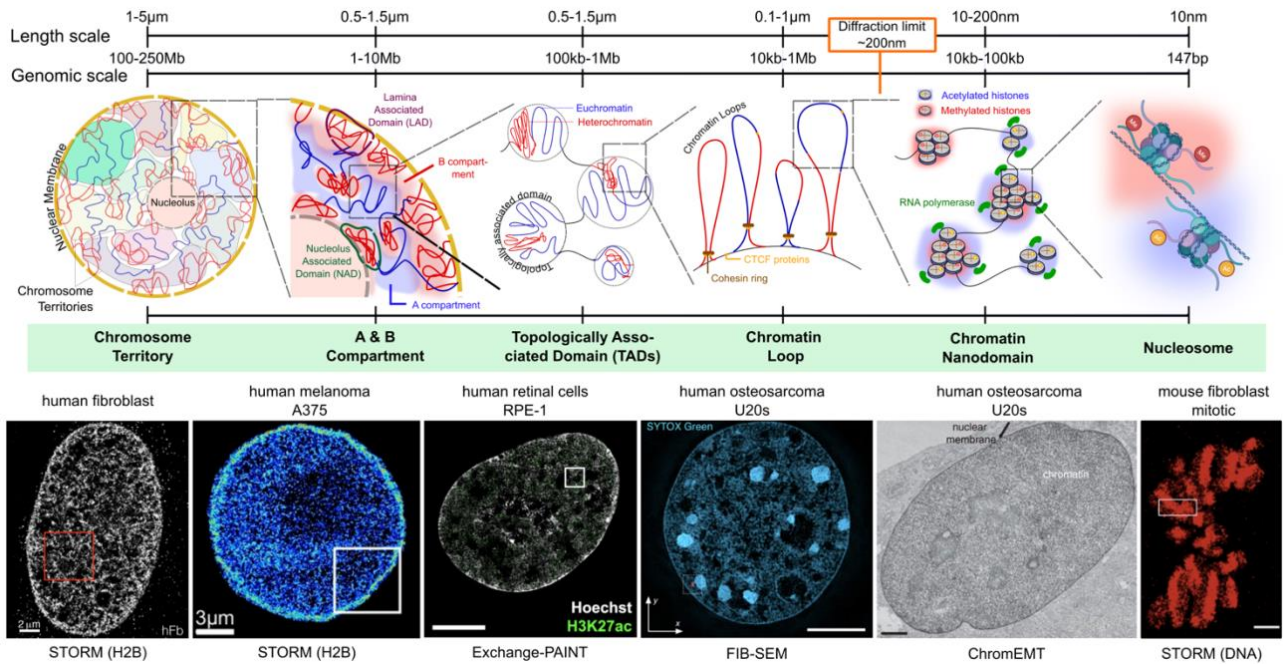

**Figure 1. Positioning Nanoscale Chromatin Domains Within Multiscale Genome Organization.** Genome organization spans a hierarchy of spatial and functional domains, from whole chromosome territories to individual nucleosomes. The schematic illustrates the key architectural levels (chromosome territories, compartments, TADs, loops, nanodomains, and nucleosomes) alongside their characteristic spatial scales. Sub-diffraction chromatin nanodomains (~50–200 nm) represent a recently uncovered layer of organization, enriched in specific histone modifications and transcriptional regulators, yet invisible to diffraction-limited microscopy. Their discovery has been enabled by super-resolution fluorescence microscopy and electron-based methods. Imaging panels demonstrate how diverse techniques converge to resolve nanoscale domains, collectively establishing their conserved and distinct structural identity[1, 2, 4, 5, 11, 12].

<u>Electron Tomography Validates Dense Chromatin Domains:</u> Electron microscopy has independently validated these structures. ChromEMT, developed by Ou et al., confirmed that chromatin in situ exists not as regular 30-nm fibers but as disordered chains of 5–24 nm fibers forming dense, irregular domains[5]. These domains varied in chromatin volume concentration (12–52%) and appeared consistent with sub-diffraction nanodomains observed by STORM. Similarly, ChromSTEM[6], with the capability to quantify chromatin mass at ~2–4 nm resolution, suggested characteristic fractal scaling and demonstrated that these structures persist after cohesin depletion, underscoring their physical rather than topological identity. Miron et al. used 3D-SIM and FIB-SEM to reveal that chromatin forms 200–300 nm domains, referred to as chromatin domains (CDs), which are arranged in chains separated by RNA-rich interchromatin space[4]. These domains exhibited a nanoscale zonation with repressive marks enriched in the interior and active transcription machinery localized to the periphery, further supporting a model of compartmentalized genome function.

<u>Imaging Genomic Loci and Epigenomic States in Context:</u> Multiplexed FISH approaches such as Oligopaint have enabled spatial mapping of hundreds to thousands of genomic loci with epigenetic and transcriptional context. Boettiger et al. showed distinct folding principles for active, inactive, and Polycomb-repressed domains using Oligopaint-STORM, revealing divergent size scaling exponents and internal mixing patterns[7]. Micro-C[33], a high-resolution Hi-C variant, has further revealed sub-TAD features that resemble nanodomains in their spatial scale and variability, especially under conditions of cohesin depletion. These findings align with recent multiplexed point-localization studies demonstrating that transcriptional and epigenetic targets stratify below ~300 nm, with histone acetylation marks (e.g., H3K27ac) forming nanoscale clusters that partition from repressive marks and DNA upon transcriptional inhibition or epigenetic drug treatment, underscoring their functional autonomy from broader compartments[15].

| Technique | Resolution | Labeling Strategy | Strengths | Limitations | Key Contributions |
|---|---|---|---|---|---|
| STORM (Ricci et al., Xu et al.) | ~20–30 nm | Antibodies (e.g., H2B, PTMs) | Revealed nanoclusters and LAD morphologies | Requires fixation; No gene level specificity | Defined nucleosome clutches; histone mark stratification |
| PALM (Nozaki et al.) | ~30–50 nm (live) | Photoactivatable fluorescent proteins | Live-cell dynamics; identifies coherent motion | Lower spatial resolution; specialized analysis | Visualized dynamic chromatin domains in live cells |
| 3D-SIM + FIB-SEM (Miron et al.) | ~100–300 nm | DNA dyes + EM stains | Correlative resolution; identifies structural zonation | Limited molecular specificity | Mapped CD chains and spatial zonation at domain surfaces |
| ChromEMT (Ou et al.) | ~8 nm | DNA-binding photooxidizing dye | High-resolution ultrastructure of chromatin | EM-based; fixed cells only | Rejected 30-nm fiber; confirmed disordered chromatin chains |
| ChromSTEM (Virk et al.) | ~2–4 nm | Electron density contrast | Density quantification; detects fractal packing | No molecular specificity | Determined fractal dimension; cohesion-independent domains |
| Oligopaint-STORM (Boettiger et al.) | ~30 nm | Locus-specific DNA probes | Folding rules by epigenetic state | Complex probe design; fixed cells | Showed scaling differences by chromatin state |
| PWS (Virk et al.) | ~20–200 nm (ensemble) | Label-free | Live-cell; chromatin packing scaling | No gene-level specificity | Linked chromatin structure to transcriptional plasticity |
| 3D-SIM + Oligopaint FISH (Szabo et al.) | ~100 nm | DNA locus-specific probes | Direct visualization of TADs and nanodomains in single cells; cohesin/CTCF perturbation tested | Fixed cells only; requires well-designed probes | Revealed nanodomains as structural units persisting without loop extrusion components |
| Exchange-PAINT + HIST (Rahman et al.) | ~22 nm lateral / ~55 nm axial | Oligo-conjugated antibodies & DNA | High multiplexing; 3D nanodomain profiling of proteins, PTMs, and DNA | Requires fixation; complex analysis | Revealed nanoscale spatial transitions of nuclear proteins below ~300 nm |

**Table 1.** Summarizing the key imaging tools to observe nanoscale domains, highlighting their advantages, limitations, and the insights they offer[1-8].

## 3. Physical and Epigenetic Architecture

Sub-diffraction chromatin domains represent not only structural units of genome organization but also epigenetically defined regulatory modules. Having established their existence, the next step is to understand their physical and chemical composition. Unlike looping driven chromatin segregation, these nanoscopic domains emerge from local nucleosome-level organization coupled to specific epigenetic states, as revealed primarily through super-resolution imaging. In the sections that follow, we consider their defining features: physical size and packing, epigenetic and transcriptional stratification, independence from loop extrusion and tethering to the nuclear and nucleolar peripheries.

Domain Size, Packing, and Fractality: Chromatin domains have been observed across imaging modalities [5, 34, 35] and across various cell lines, including, mesenchymal stem cells[9], fibroblasts[11], mouse cells[34], T cells[16], B cells[17] and various human and mouse cancer cell types[2]. A consistent feature of these domains is their characteristic size distribution, with radii typically ranging from 50–200 nm and averaging around 100 nm. The distribution is heavy-tailed, indicating the presence of larger domains up to a micron, likely corresponding to highly condensed chromatin regions such as centromeric or telomeric zones[9, 36, 37]. Additionally, their internal structure is not homogeneous; and to characterize it in accordance with previously proposed fractal globule framework of the chromatin[38], studies have fitted the observed domains against empirical power-law scaling of the chromatin mass with its radius, characterized by a fractal dimension ($D_f$) of ~2.4–2.7[6]. This scaling suggests a spatial organization more compact than a random coil ($D_f = 2$) but less dense than a globule ($D_f = 3$). Super-resolution 3D imaging and correlation-based analyses (e.g., pair cross-correlation functions) have reinforced this picture, showing evidence of a heterogenous packing behavior for DNA and associated histone marks across scales from ~50 nm to 1 µm[15]. These domains also display chromatin volume concentration gradients: denser cores tend to be less accessible and are enriched in repressive features, while lower-density peripheries are more accessible and transcriptionally engaged[35].

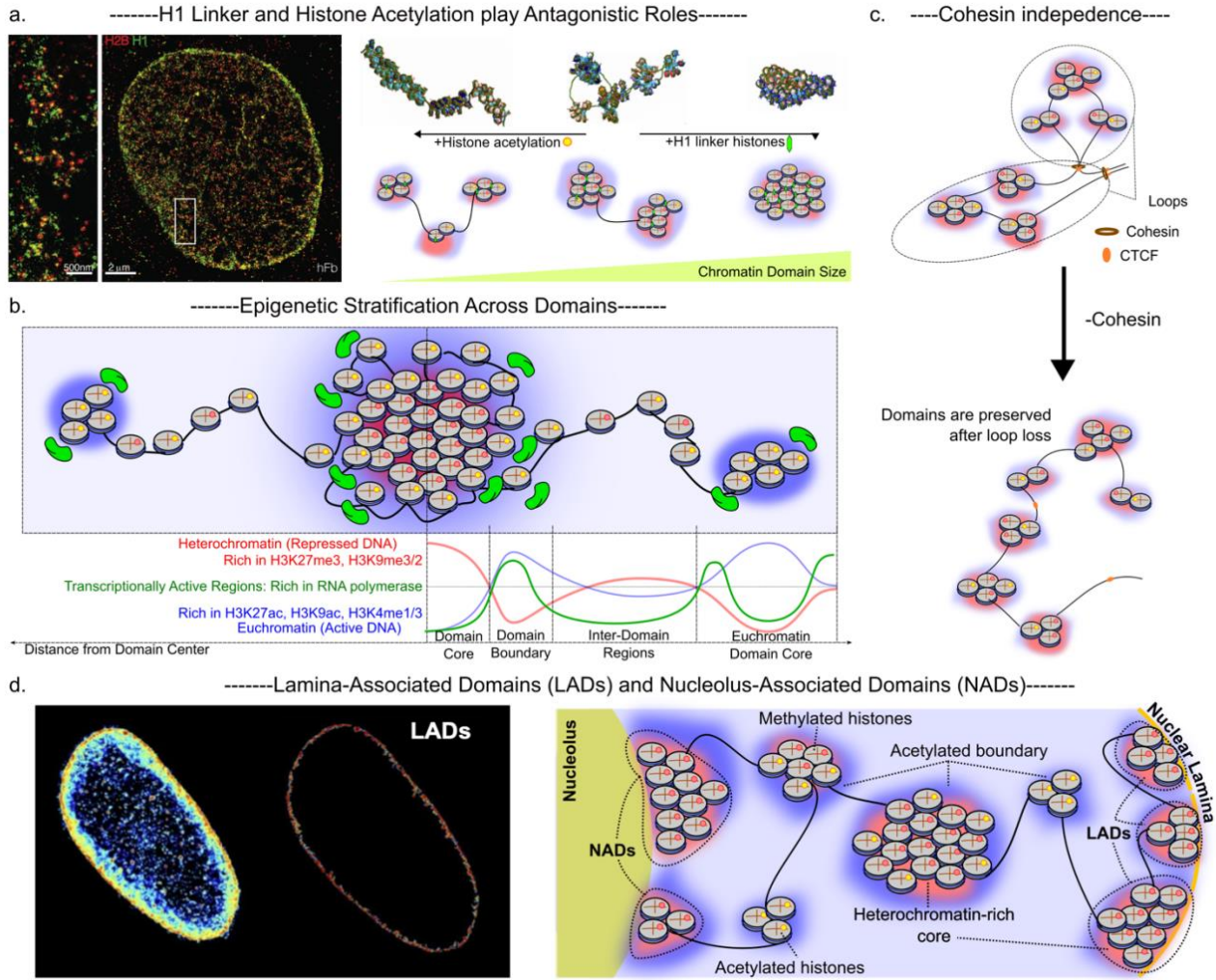

**Figure 2. Physical and epigenetic organization of sub-diffraction chromatin domains.** a) Super-resolution imaging reveals H1 linker proteins are ubiquitous to chromatin nanodomains and H1 linker proteins act as antagonists to Histone Acetylation to control the size and packing density of domains. b) These domains exhibit radial stratification of epigenetic states, with transcriptionally active modifications (for example, H3K27ac, H3K4me3) and RNA polymerase II enriched at domain boundaries, while repressive marks (H3K27me3, H3K9me3) occupy the interior. c) Absence of looping mechanisms do not affect the domains size and distribution. d. At the nuclear scale, chromatin nanodomains are tethered to architectural landmarks such as the nuclear lamina and nucleolus, forming lamina-associated domains (LADs) and nucleolus-associated domains (NADs). Together, these findings highlight that chromatin domains represent integrated physical–epigenetic modules, simultaneously defined by polymer packing principles and the spatial distribution of histone modifications.

Together, these observations establish sub-diffraction chromatin domains as structurally heterogeneous, fractal-like units whose size, packing, and internal density gradients provide a physical scaffold for differential accessibility and epigenetic regulation.

Epigenetic and Transcriptional Stratification: Nanoscale domains exhibit epigenetic heterogeneity[1] (Fig 2b). Super-resolution imaging consistently shows a spatial separation of histone marks: active modifications such as H3K27ac and H3K4me3 localize toward the edges or peripheral shells of domains, often adjacent to interchromatin space. In contrast, repressive marks like H3K27me3 and H3K9me3 are enriched in the denser core regions and often associated with larger domains, particularly at the nuclear periphery [4, 7]. The spatial positioning of these marks is functionally significant as well. Active transcription machinery (e.g., RNAPII, CDK9, p300) localizes preferentially near peripheral zones, which are rich in active chromatin markers, forming a local organization that supports rapid access and regulation[15, 39]. Repressive zones within the domain interior often colocalize with structural proteins like HP1α or components of Polycomb complexes, suggesting a role in stabilizing silenced states through compaction and phase-like behavior[15]. Because enzymes that promote heterochromatin formation are generally smaller than those that facilitate euchromatin organization, domain maintenance may be further maintained by size-dependent diffusion: smaller heterochromatin enzymes can penetrate to the core, whereas larger euchromatin enzymes are restricted to the periphery, where they support transcription[4, 40]. As we will discuss in the next section, this spatial partitioning, defined by boundary positioning, has significant biophysical underpinnings and is closely tied to transcriptional plasticity and cellular heterogeneity.

In addition to histone modifications, the distribution of linker histone H1 modulates domain architecture[1, 41] (Fig 2a). Experimental depletion of H1 results in decreased packing efficiency and increased chromatin accessibility[1]. *In silico* studies further support this, showing that H1 occupancy is a major regulator of nucleosome domain compaction, working in concert with acetylation levels and nucleosome spacing[41, 42]. Physical determinants of domain size and packing also include nucleosome spacing. The presence of nucleosome-free regions (NFRs) strongly influences domain formation. Increased NFR frequency correlates with smaller, more fragmented domains through reduced internucleosomal interactions[42]. Together, these findings establish that nanoscale domains are not passive structural aggregates but epigenetically and physically stratified modules, in which the interplay of histone marks, architectural proteins, enzyme accessibility, and nucleosome organization collectively governs compaction, accessibility, and transcriptional potential.

Structural Persistence Without Loop Extrusion: A key distinction between sub-diffraction domains and the higher order structure is its independence from the loop extrusion machinery (Fig 2c). ChromSTEM[6] has shown that these domains display a characteristic scaling behavior and are preserved after RAD21 depletion[43], indicating that their formation does not rely on loop extrusion. Similarly, Miron et al. found that the structure persists in the absence of cohesin, reinforcing the notion that these domains are physical, not just topologically extruded, units[4]. This is supported by single-cell 3D-SIM studies[8] demonstrating that nanodomains within TADs remain intact in cells depleted of either CTCF or cohesin, despite a loss of TAD-level insulation. However, this independence does not imply that loop extrusion machinery plays no role in domain size selection. Recent studies suggest that cohesin- and RNAPII-driven supercoiling can modulate the size distribution of these domains, hinting at a nuanced interplay between transcriptional activity, torsional stress, and physical compaction[37, 39], an aspect we will cover from a biophysical perspective in the next section.

Tethered Chromatin Domains at the Lamina and Nucleolus: Beyond the nuclear interior, compact chromatin domains are also observed at the nuclear periphery and around the nucleolus

(Fig 2d). Super-resolution imaging of mammalian cells reveals that lamina-associated domains (LADs), identified as discrete, heterochromatic structures exhibit an epigenetic constitution that resembles domains in the nucleus interior[9, 11, 34]. These peripheral domains vary in thickness and chromatin density depending on their epigenetic state and the presence of interacting proteins, such as HDAC3 and LAP2β[44-46]. While nucleolus-associated domains (NADs) remain less well characterized at super-resolution, their enrichment in heterochromatin and positioning near the nucleolar periphery allude that they may follow similar principles[47-49]. These findings suggest that local interactions and epigenetic states, operating through the same core biophysical principles that govern chromatin domains, could contribute to the organization of compact chromatin structures along the nuclear periphery as well.

Despite extensive imaging-based observations, sub-diffraction domains are still often treated as distinct structural entities rather than elements of a unified framework. Their physical and epigenetic features are well described, yet how these layers integrate into a coherent model of chromatin organization remains unclear. Critical questions persist, including the causal links between packing density, histone mark localization, and the functional outcomes of these universal packing principles. In the next section, we explore emerging conceptual models that aim to reconcile these findings, offering an integrative view of chromatin architecture at the nanoscale and its role in enabling cells to regulate phenotypic trajectories.

## 4. Understanding the Physics of Domain Regulation

The structural features of sub-diffraction chromatin domains, including their compact organization, radial stratification, and independence from loop extrusion, raise important questions about the molecular and physical mechanisms that govern their formation and stability. Physical modeling has been instrumental in unifying the multiple scales of chromatin organization and in developing a mechanistic understanding of their underlying principles[50-56]. Although imaging studies have characterized their morphology and compositional complexity, a mechanistic understanding of how these domains are formed and regulated and how they influence the transcriptional state of the cell remains an open challenge. Emerging theoretical and computational models, informed and validated by experimental data, suggest that these domains are not passive consequences of chromatin folding. Instead, they appear to arise from the interplay of epigenetic reaction-diffusion processes, transcriptional activity, and spatial anchoring to nuclear structures. Within this framework, chromatin domains are best viewed as steady-state attractors in a dynamic, energy-consuming landscape shaped by both biochemical and physical constraints. The underlying physical principles which have been used to understand domain size, positioning and morphology distributions are recapitulated in Box 2.

**Box 2: Physical Principles Behind Nanoscale Chromatin Domains**

Three physical principles underlie many organizing phenomena in biology, including chromatin: **diffusion and phase separation, non-equilibrium reaction,** and **wetting**. While polymer physics and protein-protein interaction energetics fundamentally shape the chromatin landscape, here we focus specifically on the distinct physical and dynamical principles that govern the formation and maintenance of chromatin domains. Each concept and their implications will be understood using simple scaling arguments.

**Diffusion and Phase Separation**

Diffusion is the tendency of molecules to spread out and erase concentration differences — the natural outcome of random thermal motion. The characteristic time for this process depends on how far a molecule must travel (distance $L$) and how fast it diffuses (diffusion coefficient $D$), with the scaling $t \sim L^2/D$. In the cell nucleus, this means that protein gradients on the order of microns disappear within just a few seconds. *A familiar analogy is the way a drop of food coloring disperses in a glass of water: even without stirring, the color gradually spreads until the liquid becomes uniform.*

When molecules additionally prefer binding to others of the same type, this balance shifts. Attractive interactions can outweigh the entropic drive to spread out, so that local fluctuations in concentration are amplified instead of erased, which is the basis of phase separation. In this regime, small clusters tend to dissolve while larger ones grow, a process known as Ostwald ripening. *This can be thought of like oil droplets in a vinaigrette: tiny droplets gradually disappear, while larger droplets merge and grow at their expense.* Left unchecked, this coarsening has no natural endpoint at equilibrium, with ever fewer and larger droplets appearing over time. Attractive nucleosome–nucleosome interactions lead to phase separation of heterochromatin and euchromatin. But phase separation alone would predict unlimited coarsening; observed nanoscale sizes indicate active arrest mechanisms.

**Non-Equilibrium Activity and Reaction-Diffusion Balance**

In an equilibrium system, molecules move randomly but there are no sustained flows; *like water sloshing in a sealed bathtub, the level stays the same over time.* Cells, however, deliberately break this balance by spending energy in the form of ATP, GTP, or metabolites. This energy allows "uphill" reactions that oppose diffusion and maintain chromatin domains in a non-equilibrium steady state. *A more useful analogy is a sink with the faucet running and the drain open: the water level looks constant, but only because continuous inflow and outflow are balanced.*

In the case of chromatin, histone-modifying enzymes consume metabolites such as acetyl-CoA and SAM to write or erase epigenetic marks, breaking detailed balance and stabilizing nanodomains at ~100 nm. Physically, diffusion delivers heterochromatin into a domain at a rate proportional to $D/R$ (where $D$ is the diffusion constant and $R$ the domain radius), while epigenetic reactions remove or remodel it at a rate proportional to $\Gamma R$ (where $\Gamma$ is the reaction rate). Balancing these fluxes fixes a steady-state size, $R_{ss} \sim \sqrt{D/\Gamma}$. In other words, energy consumption halts the indefinite coarsening expected at equilibrium and locks domains into a reproducible nanoscale size range. This same principle underlies other dynamic biological patterns: from Turing patterns on developing tissues to cytoplasmic condensates and oscillatory circuits, all of which persist only because energy is continuously consumed to sustain them.

**Wetting and Surface Interactions**

When a droplet meets a surface, its shape reflects a tug-of-war between two forces: cohesion, the attraction of molecules within the droplet to one another, and adhesion, the attraction of those molecules to the surface. This balance is described by Young's law, which relates the contact angle ($\theta$) of the droplet to the interfacial tensions between liquid, solid, and vapor. A large contact angle means the droplet beads up, cohesion dominates, and the liquid resists spreading. A small contact angle means the droplet flattens and spreads, adhesion dominates.

Everyday life provides familiar examples. *Rain on a freshly waxed car hood or on a lotus leaf forms nearly spherical beads: the water molecules cling more tightly to each other than to the water-repellent surface. In contrast, when water hits clean glass, it spreads into a thin film: here, adhesion to the glass overcomes cohesion, pulling the droplet outward.* In this way, the contact angle becomes a simple and quantitative measure of how surfaces can stabilize or reshape droplets. Within the nucleus, the nuclear lamina and nucleolus provide adhesive surfaces; tethering proteins act as anchors, turning transient condensates into lamina- or nucleolus-associated domains whose thickness is set by adhesion and reaction–diffusion balance.

Domain Emergence and Maintenance through Epigenetic Reaction-Diffusion competition: Polymer simulations[12] and theoretical analysis[9] have suggested that chromatin nanodomains form and persist through a constant tug-of-war between two fundamental dynamic processes that together define their characteristic size. First, passive diffusion drives nucleosomes carrying the same chemical tags, such as methyl groups on H3K9 or H3K27, toward one another, causing small clusters to merge into larger condensates. Such processes can be facilitated by linker proteins such as HP1 or due to electrostatic interactions[57] (Fig. 3a). Left alone, this would lead to complete phase separation by Ostwald ripening. In living cells, however, histone epigenetic read-write enzymes (for example, methyltransferases like EZH2 and deacetylases such as HDACs) continually erase and install epigenetic marks, consuming metabolic cofactors (such as acetyl-CoA) in the process[58]. By converting heterochromatin back to euchromatin inside emerging droplets, these energy-consuming reactions generate a chemical flux that counterbalances the inward diffusive flux of heterochromatin, arresting droplet growth and stabilizing domains at a characteristic nanometer scale(Fig. 3a). Because installing or removing an unfavorable mark, for example, placing a euchromatin mark within a heterochromatin-rich environment, requires energy, chromatin nanodomains should be understood as active, energy-driven structures rather than passive thermodynamic minima.

A biophysical scaling argument from these studies[36, 37] captures how these opposing fluxes set the steady-state radius $R_{ss}$ for nanodomains. As heterochromatic segments diffuse inward at a flow determined by an effective diffusion coefficient $D$, enzyme-driven heterochromatin mark removal (or euchromatin mark deposition) at rate $\Gamma_{ac}$ pushes chromatin back out. Detailed analytic derivations, confirmed by detailed phase-field[9, 36, 37] and polymer simulations[12] demonstrate that (Fig. 3b):

$$R_{ss} \propto \sqrt{\frac{D}{\Gamma_{ac}} \frac{\Gamma_{me}}{\Gamma_{me} + \Gamma_{ac}}},$$

where $\Gamma_{me}$ is the methylation rate, which accounts for repressive heterochromatin mark deposition (or euchromatin mark removal). In simpler terms, higher diffusion or methylation activity supplies more heterochromatin to the growing droplet, enlarging it, whereas greater demethylation activity removes marks more rapidly, shrinking the domain. Through experimentally validated epigenetic remodeler concentrations within the nucleus, simulations achieve nanodomain sizes of approximately 100 nm, in striking agreement with distributions exhibited by super-resolution imaging across cell lines. These findings are further supported by epigenetic drug treatments, including Trichostatin A (TSA), a broad HDAC inhibitor, and GSK343, an EZH2 inhibitor, with consistent effects observed across human and mouse mesenchymal, epithelial, and cancer cell lines[9, 11, 12].

The reaction-diffusion framework can be further extended to specify how changes to the underlying chromatin fiber architecture can effectively determine chromatin architecture. Portillo-Ledesma et al[42] simulate mesoscale chromatin "clutches" and show that local compaction is sensitive to fiber-level parameters: linker histone H1 density and nucleosome-free region (NFR) length. Their findings provide a structural basis for modulating the effective rates of methylation ($\Gamma_{me}$) and acetylation ($\Gamma_{ac}$) in reaction-diffusion models, for example, tighter clutches formed under high H1 correspond to higher effective methylation activity, while long NFRs reduce compaction exhibit the opposing effect. These structural effects can be folded into the scaling law discussed above, showing how local chromatin organization tunes global domain size. Together, these findings establish chromatin nanodomains as steady-state products of an energy-consuming reaction–diffusion balance that is further tuned by polymer-level features of the chromatin fiber. By linking enzymatic activity, metabolite availability, and nucleosome-scale organization to reproducible nanoscale structures, this framework provides a mechanistic bridge between molecular biochemistry and nuclear architecture, offering a unified physical basis for understanding how cells control mesoscale genome compaction and accessibility.

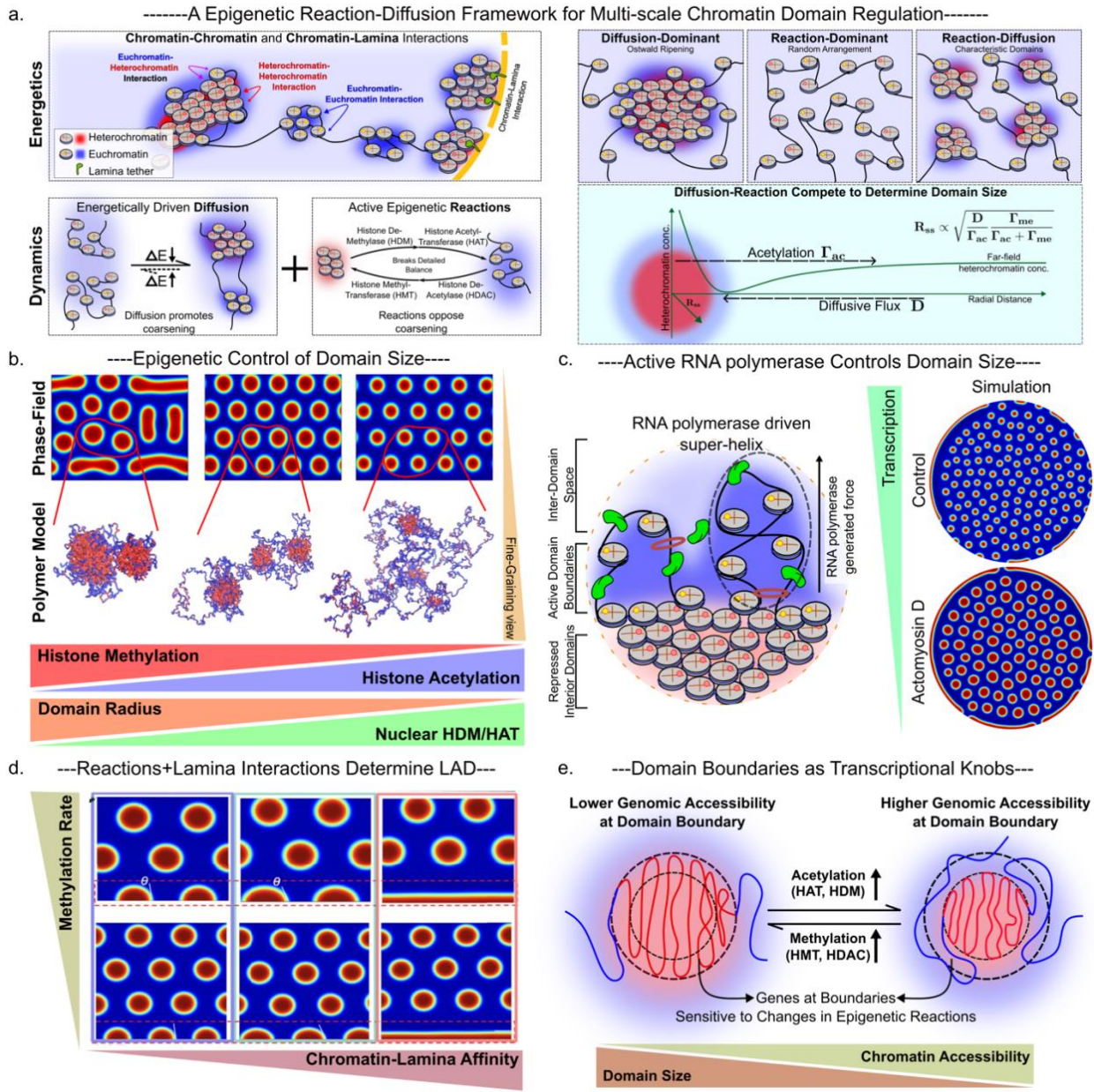

**Figure 3. A unified physical framework for the formation, regulation, and functional consequences of chromatin nanodomains.** Chromatin nanodomains arise from the interplay of diffusion, epigenetic reactions, polymer dynamics, transcriptional forces, and spatial tethering. a) Chromatin–chromatin and chromatin–lamina interactions create an energetic landscape favoring heterochromatin condensation and peripheral localization. The steady-state domain size is determined by the balance between diffusive influx and reaction-driven efflux at the domain boundary, leading to different regimes of domain size depending on relative reaction and diffusion strengths. b) Theory-driven phase field and polymer simulations capture the scaling behavior of chromatin domains with changing reactions at different scales. c) RNA polymerase II imposes mechanical flux of chromatin out of domains via supercoiling and loop extrusion, locally driving chromatin decompaction at domain boundaries. d) Chromatin–lamina affinity constrains domain positioning, leading to distinct LAD morphologies based on tethering strength (chromatin-lamina affinity) and methylation activity. The morphologies formed can thereby be characterized in terms of the LAD thickness and contact angles $\theta$. e) The domain boundary serves as a site of switchable chromatin accessibility and transcriptional regulation, where genes can dynamically transition between repressed and active states in response to changes in histone writer/eraser driven active epigenetic reaction kinetics.

Transcription-Coupled Regulation of Nanodomain Size: Domain boundaries provide a natural spatial niche for transcription, serving as sites where RNA polymerase and associated machinery are preferentially localized. Here, we turn to the role of this transcriptional machinery itself, not just as a responder to chromatin state, but as an active regulator that shapes the transcriptional landscape of the cell. Beyond the balance of diffusion and epigenetic reactions, RNA polymerase II (RNAPII) adds a mechanical layer of control: as polymerases transcribe DNA, they generate negative supercoils that drive loop extrusion at the heterochromatin-euchromatin boundary, held at the hinge by cohesin (Fig. 3c). Mechanistically, this process acts like a tug on the droplet's surface, pulling away repressive chromatin and transforming it to active chromatin. Kant et al.[37] captured this in a phase-field model by adding an extrusion-driven flux (rate $\Gamma_a$) to the previously described reaction-diffusion system. The result is a modification of the reaction-diffusion steady-state radius:

$$R_{ss} \propto \sqrt{\frac{D}{\Gamma_{ac} + \Gamma_a} \frac{\Gamma_{me}}{\Gamma_{me} + \Gamma_{ac} + \Gamma_a}}.$$

In practice, the extrusion-driven flux ($\Gamma_a$) is a function of the cohesin residence time ($\Gamma_{coh}$), and thereby its loading and unloading and the rate of transcription by RNAPII ($\Gamma_{tr}$), which act in a dependent manner, thereby $\Gamma_a = \Gamma_{coh} \times \Gamma_{tr}$. The authors validate this model through experimental data that inhibited WAPL (a cohesin unloader) and RNAPII and observed the corresponding domain shrinkage and growth. These results demonstrate that transcriptional forces act as a "tunable knob" for chromatin nanodomains across cell lines[37, 39]. This positions transcription as an active participant in shaping chromatin architecture, rather than a downstream consequence of it. By coupling RNA polymerase activity to loop extrusion, cells gain a mechanism to dynamically modulate domains, specifically at their boundaries and fine-tune gene accessibility in response to changing transcriptional demands.

Spatial Anchoring and formation of Lamina and Nucleolus Associated Domains: Peripheral chromatin domains form when heterochromatin is anchored to nuclear structures like the lamina or nucleolus through specific tethering proteins such as LAP2β, LBR, and Emerin[46] (Fig. 3a). Dhankhar et al.[36] extended the reaction–diffusion framework by incorporating chromatin–lamina adhesion, providing a physical model for the formation and shaping of LADs (Fig. 3d). In this model, LAD size emerges from the balance between chromatin diffusion, enzymatic modification rates, and the strength of tethering to the nuclear lamina. Stronger adhesion leads to broader, sheet-like LADs that spread along the nuclear periphery, whereas weaker interactions produce thinner, punctate structures (with the spreading defined through the contact angle $\theta$, explained in Box 2). These predictions are supported by super-resolution imaging, which reveals a bimodal distribution of laminar affinity, likely corresponding to low affinity near nuclear pores and high in regions enriched with anchoring proteins such as LBR and LAP2β. Importantly, the model draws from classical wetting physics: the morphology of LADs is governed by effective surface tensions between chromatin states and the nuclear lamina, much like how a droplet spreads on a solid surface. The contact angle $\theta$ in this analogy reflects the energetic favorability of peripheral chromatin association. Thus, LADs are not rigid compartments but dynamic structures whose size and shape adapt to biochemical and mechanical cues. Consistent with this view, live-cell super-resolution imaging shows that tethering not only affects the spreading behavior of LADs but also constrains their motion, linking spatial anchoring to the regulation of chromatin mobility and nuclear organization[3].

By analogy, nucleolus-associated domains (NADs) form via tethers like nucleolin and NPM1 that bind repressive chromatin to the nucleolar surface[59] (Fig. 3a). Although super-resolution data on NADs remain less analyzed, natural extensions of current theoretical models predict that adding a similar adhesion term for nucleolar tethers recapitulates the heterochromatin clusters observed around the nucleolus. Mechanistically, both LADs and NADs demonstrate how active epigenetic reactions, such as methylation and acetylation, generate nanodomains whose size, shape, and peripheral positioning are stabilized by specific protein–membrane interactions. In this way, non-equilibrium events are converted into stable, transcriptionally silent compartments. This principle is particularly significant given that nearly 40% of chromatin is associated with the nuclear periphery, underscoring its central role in large-scale gene silencing[60].

Domain Boundaries as Sites of Epigenetic Remodeling and Transcriptional Control: A key feature of reaction-diffusion models is that chromatin nanodomains develop non-uniform internal structure, with biochemical activity localized at their edges. As repressive marks like H3K9me3 accumulate through inward diffusion and nucleosomal attraction, a densely packed core forms that is energetically stable and largely inaccessible to enzymatic remodeling. Surrounding this core, however, is a depletion layer: a peripheral region where chromatin is less compact, physically accessible and most acetylated (Fig. 2b, 3a). Simulations show that this layer naturally emerges from the coupling of polymer mechanics and reaction kinetics and serves as the primary site of epigenetic turnover. Because modifying the core would require breaking favorable heterochromatin-heterochromatin interactions, mark removal reactions, such as acetylation or demethylation, occur majorly at the boundary, where the energetic cost is lower and enzymes can more easily access the chromatin fiber. This creates a localized reaction zone that counteracts the inward flux of heterochromatin, stabilizing domain size and composition. The model thus predicts[12, 36, 37], and imaging confirms[4], that euchromatic marks and transcriptional regulators are enriched at domain boundaries, where physical accessibility and epigenetic responsiveness converge.

This boundary-localized remodeling has important implications for transcriptional plasticity. Because epigenetic reaction driven gene accessibility alterations are concentrated at domain edges, genes positioned near these boundaries are uniquely poised to respond to fluctuations in reaction rates. Simulations show that changes in the rates of methylation or acetylation—whether driven by shifts in enhanced expression of epigenetic remodelers, changes to the cell metabolic state, or through external chemo-mechanical signaling leading to shuttling of enzymes to the nucleus—can lead to rapid and localized changes in domain size, disproportionately affecting chromatin at the periphery. As the boundary moves outward or inward, genes near this interface can be brought into or pushed out of a repressive environment without requiring global chromatin reorganization. This creates a highly tunable mechanism for controlling gene accessibility: boundary-proximal genes can switch transcriptional states quickly, while those buried deep within domains remain insulated from transient fluctuations (Fig. 3e). Such spatial compartmentalization of responsiveness enables the cell to maintain stable gene silencing at the core while preserving reactive capacity at the edge, offering a physical basis for gene silencing though compaction and rapid epigenetic adaptation in response to developmental cues or environmental stress. In the next section, we will discuss how such localized regulation manifests in desirable ways in pluripotency and undesirable ways in cancer. Although both states exhibit high transcriptional plasticity, arising from smaller domain cores and increased effective surface area, pluripotent cells use this flexibility to pursue multiple developmental trajectories, whereas cancer cells exploit it to withstand stress and acquire drug tolerance.

Concludingly, sub-diffraction chromatin domains emerge from an active balance of epigenetic reactions, transcriptional forces, and nuclear tethering. Reaction–diffusion models provide a mechanistic framework for how domain growth is arrested, transcription modulates domain size, and anchoring fixes spatial position within the nucleus. In parallel, alternative approaches based on self-returning polymer dynamics[61-63] have also recapitulated observed domain geometries and hypothesized fractal-like scaling behavior, but they omit essential factors such as non-equilibrium enzymatic activity and environmental responsiveness. These models offer a valuable geometric reference but overlook the dynamic, regulated nature of chromatin domains in living cells. A major outstanding question is whether the proposed fractal-like scaling of these domains[6, 15] is truly observed at multiple scales and to understand their causality. Resolving this will require multiscale, high-fidelity imaging combined with perturbative experiments and integrative modeling to bridge physical structure with biological function.

## 5. Functionality and Responsiveness of the Domains through Development, Disease and Extracellular Cues

The previous section outlined a working model in which active reactions, driven by epigenetic write/erase enzymes (such as HDACs, HMTs, HATs and HDMs) define the global size distribution of the sub-diffraction chromatin domains. Here we examine how this cascade affects the phenotypic trajectories of the cell. As established, through computational and theoretical modeling, rather than static bundles of DNA, chromatin domains can act as modular control units: the outer shell favors transcription, the inner core maintains repression, and the balance between the two shifts signals (reflected through the epigenetic reaction rates). Such shifts can occur when the cell encounters changes through processes such as extracellular mechanical stress, metabolic change or developmental signals. The case studies that follow—spanning pluripotency, immune activation, extracellular cues, and cancer—illustrate how alterations in domain size distribution, particularly boundary driven transcription control, translates into changes in gene expression and cell identity. They also highlight where the current framework succeeds and where new mechanistic are needed.

Domain Distribution Response to Extracellular Mechanical and Chemical Cues: Microenvironmental cues such as substrate stiffness, oxygen tension, and inflammatory signals modulate chromatin domain architecture by altering epigenetic writer/eraser concentrations in the nucleus[9]. Super-resolution imaging reveals that softer environments increase domain sizes in hMSCs, possibly through H3K27me3 driven epigenetic regulation also leading to peripheral condensation i.e., increase of lamina-associated domains (LADs) (Fig. 4a). Hypoxic stress also leads to an increase in the domain size. Quantitatively, soft 3 kPa matrices or 1 % $O_2$ increase mean LAD thickness from 70 nm to 150 nm and enlarge interior clusters by ~40 %. These changes are hypothesized to arise through cytoplasm-to-nucleus shuttling of epigenetic remodelers like HDACs and EZH2 (leading to an effective increase in methylation rates $\Gamma_{me}$), a mechanism that may serve as a general conduit linking external inputs to transcriptional control[64, 65]. Such shifts in modification rates reshape domain size distributions and spatially reorganize open and closed chromatin, thereby altering transcriptional potential and cell identity.

Tendinosis imposes a distinct nuclear phenotype that differs from normal aging (Fig. 4a). In degenerative tenocytes, chromatin relocalizes to the periphery and condenses into thickened LADs, while interior domains remain coarse even when stiffening cues are applied[9]. Aged cells, in contrast, retain partial decondensation capacity, indicating that degeneration and aging produce divergent architectural outcomes. Within the reaction-diffusion framework, tendinosis reflects trapping in a high-$\Gamma_{me}$, high lamina-affinity state, whereas healthy and aged cells can still shift domain distributions. This loss of mechanoresponsive plasticity distinguishes chronic degeneration from gradual aging and positions domain remodeling as a central mechanism in tendon disease.

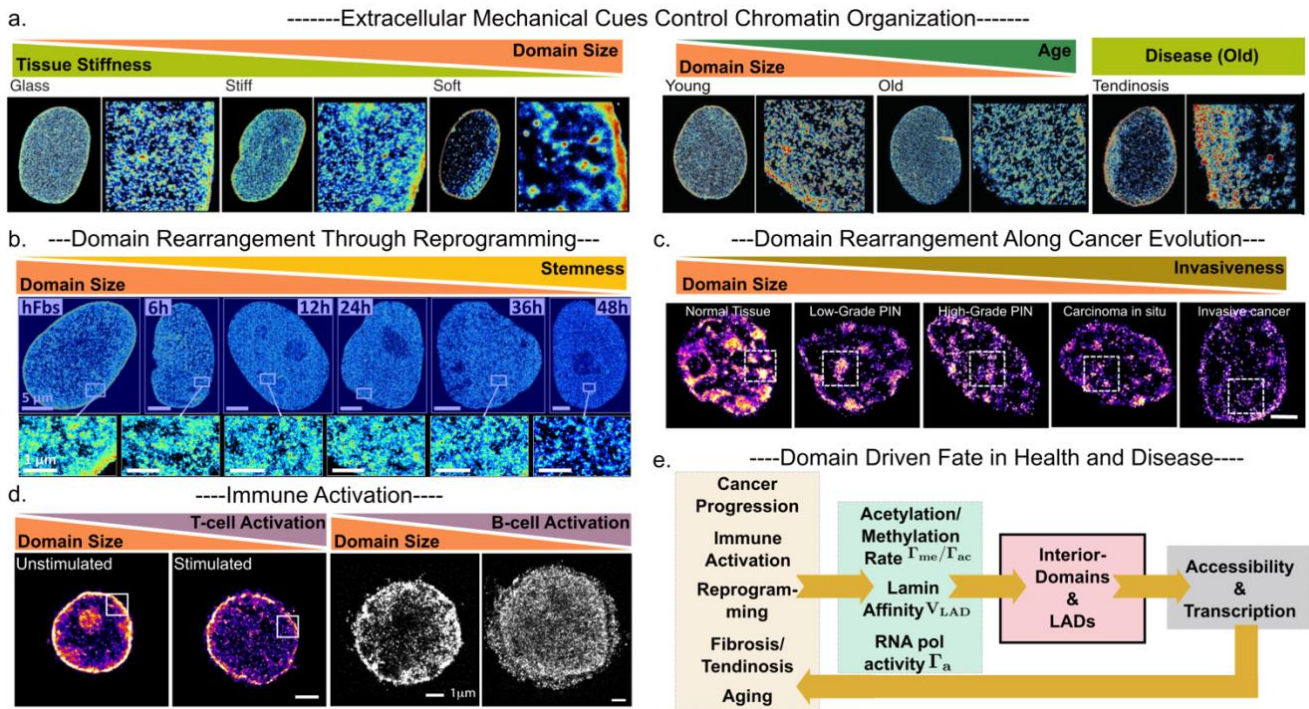

**Figure 4. Chromatin domain remodeling as a unifying principle across biological contexts.** Super-resolution imaging reveals that domain size and boundary organization change systematically in response to diverse physiological and pathological cues. a) Human mesenchymal stem cells on stiff substrates show smaller, more dispersed domains, whereas soft matrices or hypoxia promote compaction and thickened lamina-associated domains (LADs)[9]. Aging also leads to changes in condensation where domain sizes decrease with age, while tendinosis produces a distinct nuclear phenotype with persistent peripheral LADs and coarse interior domains, highlighting the divergence between degeneration and aging[9]. b) During heterokaryon-driven reprogramming of human fibroblasts, domain remodeling unfolds over days, with progressive chromatin decompaction and fragmentation of repressive cores restoring pluripotency[13]. c) In cancer progression, domains gradually decompact and fragment, transitioning from normal tissue through premalignant lesions to invasive cancer, with early loss of compact cores and boundary integrity[14, 15]. d) Immune activation exploits domain plasticity, with both T- and B-cell stimulation triggering domain fragmentation, rim broadening, and expansion of transcriptionally active nuclear space[15-17]. e) Schematic These diverse outcomes can be unified under a reaction–diffusion feedback framework, where shifts in methylation/acetylation balance ($\Gamma_{me}/\Gamma_{ac}$), lamina affinity ($V_{LAD}$), and RNA polymerase activity ($\Gamma_a$) regulate domain size and LAD organization, thereby controlling gene accessibility and transcriptional output.

Evolution of Domains through Cellular Reprogramming: During developmental transitions such as differentiation and reprogramming, chromatin domains undergo pronounced reorganization that both reflects and enables shifts in transcriptional identity. Embryonic stem cells harbor small, diffuse domains with low compaction, enriched in active histone marks[6, 34]. As differentiation proceeds, domains coalesce into larger, denser structures driven by rising linker histone H1 level, promoting heterochromatin formation and restricting transcriptional plasticity. Within the theoretical framework this increases the effective methylation to acetylation ratio ($\Gamma_{me}/\Gamma_{ac}$), contracts the accessible chromatin and yields larger, denser domains that restrict transcriptional plasticity in a boundary driven manner. Reprogramming to induced pluripotency reverses these changes: repressive cores fragment, active rims expand, and chromatin becomes more accessible at key loci such as *Nanog*[1, 66] (Fig. 4b). Martínez-Sarmiento *et al.* showed that the Nanog-associated domain decompacts before Nanog transcription increases, underscoring the regulatory importance of local chromatin domain architecture. Consistent with this principle, super-resolution imaging of heterokaryons reveals a gene-specific sequence of events: Increase in global acetylation leading to changes at the Nanog loci, which first undergoes local domain opening and later displays enhanced expression [13]. These findings confirm that chromatin domain remodeling acts at both gene-specific and global scales, regulating pluripotency-associated loci as well as broader transcriptional programs during reprogramming.

Role of Domains in Cell Transition to Cancer: Nanoscale chromatin remodeling is emerging as a universal hallmark of cancer, characterized by progressive decompaction and redistribution of higher-order chromatin structures (Fig. 4c). Super-resolution imaging in both human and mouse tissues reveals that during early carcinogenesis, even in phenotypically normal cells, domains gradually lose their compact cores, leading to global chromatin decompaction and segregated boundaries, transitioning toward fragmented nanoclusters with reduced H3K9me3 and increased euchromatic signatures[67, 68]. In the reaction-diffusion framework this shift implies a sustained rise in the euchromatin-favoring reaction rate relative to the condensation rate ($\Gamma_{ac}/\Gamma_{me}$), driving domains toward the high-plasticity region of parameter space. This evolution in domain topology is conserved across tumor types including colorectal, prostate, and pancreatic cancers, and proceeds in a stepwise fashion from precancerous lesions to invasive tumors. In ovarian cancer, cancer stem cells (CSCs) exhibit increased numbers of poised and transcriptionally active domains, smaller clutch size, which correlate with greater transcriptional plasticity and chemotherapy resistance[14, 69]. Notably, chromatin domain decompaction often precedes overt changes in gene expression or nuclear morphology, suggesting that nanoscale architecture serves as an early enabler of transcriptional deregulation and genomic instability[67]. Together, these findings position chromatin domain disruption as a globally conserved and functionally significant process that underlies the plasticity and progression of malignant states.

T and B-cell activation: Immune stimulation exploits chromatin-domain plasticity to deliver rapid transcriptional alterations. In B cells, antigen engagement elevates Myc and acetyl-CoA synthesis, raising the histone-acetylation rate ($\Gamma_{ac}$). Super-resolution studies show that peripheral heterochromatin nanodomains fragment, nucleoplasmic active space expands and enhancer–promoter loops proliferate, supporting the surge in immunoglobulin transcripts[17] (Fig. 4d). T-cell receptor signaling produces a similar response: chromatin decondenses genome-wide, median nanodomain diameter falls by roughly 40%, H3K4me3-rich rims, corresponding acetylated domain boundaries, broaden and large RNA-polymerase assemblies accumulate, coincident with cytokine induction[16] (Fig. 4d). Within the biophysical framework of Section 4, both lineages undergo a transient shift toward the high-$\Gamma_{ac}$, low-compaction corner of phase space, where smaller domains allow increased accessibility and maximize transcriptional plasticity. These examples underscore that regulated changes in domain size, compaction and position form a conserved mechanism for state-specific gene activation in the immune system.

Together, these findings point to a unifying cascade that connects cell state with chromatin nano-architecture (Fig. 4e). Mechanical, metabolic, and developmental cues alter epigenetic reaction rates, lamin affinity and transcriptional activity, which in turn reshape nanodomain distributions and the balance between active rims and repressive cores. When cells require plasticity, as in pluripotency, immune activation or cancer, domains become smaller, more open and less compacted, enabling broad transcriptional responsiveness. In contrast, differentiated or some degenerative states like tendinosis are marked by enlarged domains, thicker LADs and stronger compaction, restricting accessibility. This simple framework suggests that chromatin domains act as universal regulatory units that translate molecular kinetics and physical compaction into transcriptional potential and determine whether cells remain plastic or committed.

## 6. Conclusion and Perspectives

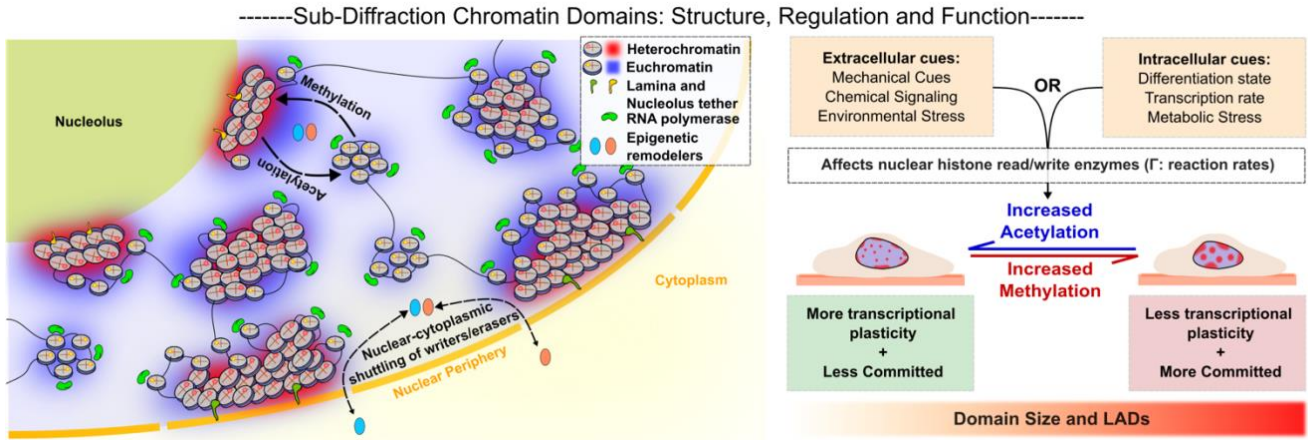

**Figure 5. Sub-diffraction chromatin domains as integrators of nuclear organization and cell fate.** Chromatin is organized into nanoscale domains composed of transcriptionally active rims and repressive cores, stabilized through the interplay of epigenetic read/write enzymes, transcriptional forces, and nuclear tethering. These domains act as steady-state attractors in an active chromatin landscape, where diffusion-driven clustering is balanced by enzymatic turnover of histone marks. Intracellular factors (such as transcription rate, differentiation state, and metabolic stress) and extracellular inputs (including mechanical cues, signaling, and environmental stress) regulate nuclear availability of epigenetic remodelers, thereby shifting reaction rates that tune domain size and lamina association. The balance between acetylation and methylation determines whether domains favor transcriptional plasticity and responsiveness, or compaction and commitment. Through this dynamic regulation, nanoscale domains serve as modular control units that couple environmental and intracellular cues to transcriptional programs, epigenetic memory, and ultimately, cell identity.

The body of evidence reviewed here positions sub-diffraction chromatin domains—clutches, nanodomains, or packing domains—as fundamental building blocks of nuclear organization. Their discovery through super-resolution and correlative electron microscopy revealed compact, heterogeneous chromatin clusters well below the diffraction limit, conserved across mammalian cell types and species. Detailed characterization has since established their defining properties: radially stratified organization of histone marks, transcriptionally active rims enriched with RNA polymerase II and cofactors, repressive cores stabilized by linker histones and silencing complexes, and structural persistence even in the absence of cohesin or CTCF. These nanoscale units stand apart from TADs and loops, capturing features of genome folding that escape ensemble-averaged contact maps and diffraction-limited imaging.

Beyond their descriptive characterization, a central outcome of the work reviewed here is the realization that the diverse processes shaping sub-diffraction domains are fundamentally unified within an active chromatin landscape (Fig. 5). Enzymatic read/write reactions, diffusion-driven clustering, transcription-induced supercoiling, and tethering to nuclear lamina (and nucleoli) come together to converge on the fundamental principle: chromatin domains are stabilized as steady-state attractors of out-of-equilibrium tug-of-wars. Reaction-diffusion models provide the most direct articulation of this view, explaining how antagonistic fluxes of mark deposition and erasure arrest coarsening at nanoscale radii. These frameworks, when coupled to multiscale formulations such as phase-field and polymer simulations, not only reproduce the experimentally observed size distributions and radial stratification but also predict how domain remodeling responds to epigenetic challenges, transcriptional disruption, or mechanical perturbation. In this sense, biophysical theory serves as a unifying scaffold that integrates imaging, functional assays, and perturbative experiments into a coherent, first-principles model of genome regulation.

Across diverse contexts—development, immune activation, mechanical stress, tissue degeneration, aging and cancer—a consistent principle emerges; chromatin domains and their boundaries act as the nanoscale switches that set cellular trajectories. The core–rim stratification ensures that transcriptional activity is concentrated at boundaries, and modest shifts in epigenetic write/erase balance are sufficient to change rim positioning and thereby alter gene accessibility and epigenetic priming. Increased methylation and higher lamina tethering promotes compaction, locking cells into repressive or committed states as seen in tendinosis or differentiation; increased acetylation expands rims and fragments cores, enhancing plasticity as in pluripotency, immune activation, or carcinogenesis (Fig. 5). In this sense, the fundamental biochemical processes of mark deposition and removal are amplified by the physics of domain organization into global control over plasticity and commitment. Boundaries therefore serve as the key integration points where non-equilibrium reaction–diffusion kinetics are converted into large-scale transcriptional malleability and phenotypic outcomes, making sub-diffraction domains the natural regulatory units through which cells navigate between stability and flexibility in phenotype.

Despite major advances, important questions remain to unpack the role and relevance sub-diffraction chromatin domains in the unified framework of genome regulation. It remains to be determined how widespread these domains are across species and whether they serve an evolutionarily conserved function. The fractal-like scaling behavior[15, 61], hypothesized across imaging modalities, remains mechanistically unresolved, raising the question of whether it reflects a regulated steady state or an imaging artifact. Related to this is the observation that not all domains are spherical: elongated structures and lamellar morphologies have been seen, suggesting that domain shape itself needs deeper investigation. Equally underexplored is the dynamic dimension: timescales of domain reconfiguration, nucleation of domains and transcriptional bursting are only beginning to be measured, yet these kinetic features are central to linking nanoscale architecture with regulatory function. A further challenge lies in reconciling perspectives: imaging has richly described nanoscale structure, but integration with sequencing-based data at the scale of domains is still lacking, limiting our ability to connect physical domains with genome-wide activity. Expanding the palette of perturbations to understand the causality cascade is also critical. While histone-modifying drugs have been widely used, the effects of osmotic stress, compressive load, shear, and tensile forces on the nucleus remain poorly characterized, even though such forces are intrinsic to development and metastasis progression[70, 71]. Additionally, the role of other important DNA perturbations, such as DNA damage[72-74] or topoisomerase action[75, 76] at the nanoscale in understudied. Finally, although suggestive evidence points to central roles in epigenetic memory[77] and cellular heterogeneity[78], the logic for whether and how nanoscale organization encodes history and produces divergent cell fates is still elusive.

Theoretical approaches will be essential in bridging these gaps. Mechanical models that link chromatin stiffness, nuclear viscoelasticity, and large-scale deformations are poised to clarify how physical stresses reshape domain architecture[79]. Biophysical modeling offers the natural route to disentangle timescales of transcriptional dynamics and epigenetic modification, providing a kinetic backbone to complement static structural data. At the same time, the field must move beyond coarse-grained representations: fine-grained and even atomistic simulations are increasingly necessary to capture the molecular detail underpinning histone modifications, nucleosome interactions, and polymer folding[80]. Because domain remodeling is already implicated in cancer, stem-cell plasticity, and degenerative disease, refining these models carries direct translational weight, which is already being explored[81]: only by understanding the mechanistic underpinnings can we hope to harness nanoscale chromatin organization for therapeutic purposes.

The coming decade promises to be transformative, as reliable, higher-resolution experimental and informed computational advances converge on the precision of nanoscale chromatin architecture. Multiscale imaging, integrated with genome-wide spatial transcriptomics, will allow direct mapping of how domain-level remodeling propagates to transcriptional programs at the single-cell scale. Chemo-mechanical perturbative approaches will reveal how nuclear architecture integrates environmental cues, while synthetic interventions such as CRISPR-based control of epigenetic marks[82] or targeted tethering molecules offer a route to reprogram domain states in real time. Equally important, interpretable machine learning and deep learning frameworks, exemplified by models such as O-SNAP and AINU[83-85] are beginning to discriminate nanoscale patterns invisible to human observers and hold the promise of disentangling causal from correlative features in chromatin organization. Together, these approaches point to a future in which structural, biochemical, and computational lenses are seamlessly combined to track domain behavior in live single cells.

A central goal will be to move toward a unified definition of *cell state*: one that merges nanoscale chromatin architecture with transcriptional and epigenomic profiles into a coherent, measurable entity. Establishing this genomic–physical picture at single-cell resolution will provide the substrate for building *in silico* models of cellular evolution, with direct applications to dormancy, drug resistance, neurodegeneration, and aging. Sub-diffraction chromatin domains lie at the heart of this vision: as the modular units through which the genome encodes responsiveness, memory, and plasticity, they represent the natural scale at which cellular identity is both stored and reshaped. By uniting molecular biology, high-resolution imaging and single-cell sequencing using physically grounded theory and artificial intelligence the field is poised not only to understand how domains structure the genome across space and time but also to harness this architecture to predict, and ultimately reprogram, cellular trajectories.

**References:**

1. Ricci, M.A., et al., *Chromatin fibers are formed by heterogeneous groups of nucleosomes in vivo.* Cell, 2015. **160**(6): p. 1145-58.
2. Xu, J., et al., *Super-Resolution Imaging of Higher-Order Chromatin Structures at Different Epigenomic States in Single Mammalian Cells.* Cell Rep, 2018. **24**(4): p. 873-882.
3. Nozaki, T., et al., *Dynamic Organization of Chromatin Domains Revealed by Super-Resolution Live-Cell Imaging.* Mol Cell, 2017. **67**(2): p. 282-293 e7.
4. Miron, E., et al., *Chromatin arranges in chains of mesoscale domains with nanoscale functional topography independent of cohesin.* Sci Adv, 2020. **6**(39).
5. Ou, H.D., et al., *ChromEMT: Visualizing 3D chromatin structure and compaction in interphase and mitotic cells.* Science, 2017. **357**(6349).
6. Virk, R.K.A., et al., *Disordered chromatin packing regulates phenotypic plasticity.* Sci Adv, 2020. **6**(2): p. eaax6232.
7. Boettiger, A.N., et al., *Super-resolution imaging reveals distinct chromatin folding for different epigenetic states.* Nature, 2016. **529**(7586): p. 418-22.
8. Szabo, Q., et al., *Regulation of single-cell genome organization into TADs and chromatin nanodomains.* Nat Genet, 2020. **52**(11): p. 1151-1157.
9. Heo, S.J., et al., *Aberrant chromatin reorganization in cells from diseased fibrous connective tissue in response to altered chemomechanical cues.* Nat Biomed Eng, 2023. **7**(2): p. 177-191.
10. Lelek, M., et al., *Single-molecule localization microscopy.* Nat Rev Methods Primers, 2021. **1**.
11. Otterstrom, J., et al., *Super-resolution microscopy reveals how histone tail acetylation affects DNA compaction within nucleosomes in vivo.* Nucleic Acids Res, 2019. **47**(16): p. 8470-8484.

12. Vinayak, V., et al., *Polymer model integrates imaging and sequencing to reveal how nanoscale heterochromatin domains influence gene expression.* Nat Commun, 2025. **16**(1): p. 3816.
13. Martinez-Sarmiento, J.A., M.P. Cosma, and M. Lakadamyali, *Dissecting gene activation and chromatin remodeling dynamics in single human cells undergoing reprogramming.* Cell Rep, 2024. **43**(5): p. 114170.
14. Xu, J., et al., *Super-resolution imaging reveals the evolution of higher-order chromatin folding in early carcinogenesis.* Nat Commun, 2020. **11**(1): p. 1899.
15. Rahman, F., et al., *Mapping the nuclear landscape with multiplexed super-resolution fluorescence microscopy.* Nature Communications, 2025. **16**(1): p. 6042.
16. Xu, J., et al., *Super-resolution imaging of T lymphocyte activation reveals chromatin decondensation and disrupted nuclear envelope.* Commun Biol, 2024. **7**(1): p. 717.
17. Kieffer-Kwon, K.R., et al., *Myc Regulates Chromatin Decompaction and Nuclear Architecture during B Cell Activation.* Mol Cell, 2017. **67**(4): p. 566-578 e10.
18. Rust, M.J., M. Bates, and X. Zhuang, *Sub-diffraction-limit imaging by stochastic optical reconstruction microscopy (STORM).* Nat Methods, 2006. **3**(10): p. 793-5.
19. Betzig, E., et al., *Imaging intracellular fluorescent proteins at nanometer resolution.* Science, 2006. **313**(5793): p. 1642-5.
20. Jerkovic, I. and G. Cavalli, *Understanding 3D genome organization by multidisciplinary methods.* Nat Rev Mol Cell Biol, 2021. **22**(8): p. 511-528.
21. Li, H., et al., *Chromosome compartmentalization: causes, changes, consequences, and conundrums.* Trends Cell Biol, 2024.
22. Bannister, A.J. and T. Kouzarides, *Regulation of chromatin by histone modifications.* Cell Res, 2011. **21**(3): p. 381-95.
23. Wang, S., et al., *Spatial organization of chromatin domains and compartments in single chromosomes.* Science, 2016. **353**(6299): p. 598-602.
24. McCord, R.P., N. Kaplan, and L. Giorgetti, *Chromosome Conformation Capture and Beyond: Toward an Integrative View of Chromosome Structure and Function.* Mol Cell, 2020. **77**(4): p. 688-708.
25. Burgess, D.J., *Chromosome structure at micro-scale.* Nat Rev Genet, 2020. **21**(6): p. 337.
26. Schnitzbauer, J., et al., *Super-resolution microscopy with DNA-PAINT.* Nat Protoc, 2017. **12**(6): p. 1198-1228.
27. Fudenberg, G. and L.A. Mirny, *Higher-order chromatin structure: bridging physics and biology.* Curr Opin Genet Dev, 2012. **22**(2): p. 115-24.
28. Banigan, E.J. and L.A. Mirny, *Loop extrusion: theory meets single-molecule experiments.* Curr Opin Cell Biol, 2020. **64**: p. 124-138.
29. Prakash, K., et al., *Superresolution imaging reveals structurally distinct periodic patterns of chromatin along pachytene chromosomes.* Proc Natl Acad Sci U S A, 2015. **112**(47): p. 14635-40.
30. Sati, S. and G. Cavalli, *Chromosome conformation capture technologies and their impact in understanding genome function.* Chromosoma, 2017. **126**(1): p. 33-44.
31. Maeshima, K., S. Ide, and M. Babokhov, *Dynamic chromatin organization without the 30-nm fiber.* Curr Opin Cell Biol, 2019. **58**: p. 95-104.

32. Lakadamyali, M. and M.P. Cosma, *Visualizing the genome in high resolution challenges our textbook understanding.* Nat Methods, 2020. **17**(4): p. 371-379.
33. Krietenstein, N., et al., *Ultrastructural Details of Mammalian Chromosome Architecture.* Mol Cell, 2020. **78**(3): p. 554-565 e7.
34. Ricci, M.A., M.P. Cosma, and M. Lakadamyali, *Super resolution imaging of chromatin in pluripotency, differentiation, and reprogramming.* Curr Opin Genet Dev, 2017. **46**: p. 186-193.
35. Li, Y., et al., *Analysis of three-dimensional chromatin packing domains by chromatin scanning transmission electron microscopy (ChromSTEM).* Sci Rep, 2022. **12**(1): p. 12198.
36. Dhankhar, M., et al., *Revealing the biophysics of lamina-associated domain formation by integrating theoretical modeling and high-resolution imaging.* Nat Commun, 2025. **16**(1): p. 7909.
37. Kant, A., et al., *Active transcription and epigenetic reactions synergistically regulate meso-scale genomic organization.* Nat Commun, 2024. **15**(1): p. 4338.
38. Mirny, L.A., *The fractal globule as a model of chromatin architecture in the cell.* Chromosome Res, 2011. **19**(1): p. 37-51.
39. Neguembor, M.V., et al., *Transcription-mediated supercoiling regulates genome folding and loop formation.* Mol Cell, 2021. **81**(15): p. 3065-3081 e12.
40. Maeshima, K., et al., *The physical size of transcription factors is key to transcriptional regulation in chromatin domains.* J Phys Condens Matter, 2015. **27**(6): p. 064116.
41. Gomez-Garcia, P.A., et al., *Mesoscale Modeling and Single-Nucleosome Tracking Reveal Remodeling of Clutch Folding and Dynamics in Stem Cell Differentiation.* Cell Rep, 2021. **34**(2): p. 108614.
42. Portillo-Ledesma, S., et al., *Nucleosome Clutches are Regulated by Chromatin Internal Parameters.* J Mol Biol, 2021. **433**(6): p. 166701.
43. Li, W.S., et al., *Chromatin packing domains persist after RAD21 depletion in 3D.* bioRxiv, 2024: p. 2024.03.02.582972.
44. Somech, R., et al., *The nuclear-envelope protein and transcriptional repressor LAP2beta interacts with HDAC3 at the nuclear periphery, and induces histone H4 deacetylation.* J Cell Sci, 2005. **118**(Pt 17): p. 4017-25.
45. Harr, J.C., A. Gonzalez-Sandoval, and S.M. Gasser, *Histones and histone modifications in perinuclear chromatin anchoring: from yeast to man.* EMBO Rep, 2016. **17**(2): p. 139-55.
46. Manzo, S.G., L. Dauban, and B. van Steensel, *Lamina-associated domains: Tethers and looseners.* Curr Opin Cell Biol, 2022. **74**: p. 80-87.
47. Padeken, J. and P. Heun, *Nucleolus and nuclear periphery: velcro for heterochromatin.* Curr Opin Cell Biol, 2014. **28**: p. 54-60.
48. Bersaglieri, C. and R. Santoro, *Genome Organization in and around the Nucleolus.* Cells, 2019. **8**(6).
49. Peng, T., et al., *Mapping nucleolus-associated chromatin interactions using nucleolus Hi-C reveals pattern of heterochromatin interactions.* Nat Commun, 2023. **14**(1): p. 350.
50. Jost, D., et al., *Modeling epigenome folding: formation and dynamics of topologically associated chromatin domains.* Nucleic Acids Res, 2014. **42**(15): p. 9553-61.

51. Abdulla, A.Z., C. Vaillant, and D. Jost, *Painters in chromatin: a unified quantitative framework to systematically characterize epigenome regulation and memory.* Nucleic Acids Res, 2022. **50**(16): p. 9083-9104.
52. Fudenberg, G., et al., *Formation of Chromosomal Domains by Loop Extrusion.* Cell Rep, 2016. **15**(9): p. 2038-49.
53. Owen, J.A., D. Osmanovic, and L. Mirny, *Design principles of 3D epigenetic memory systems.* Science, 2023. **382**(6672): p. eadg3053.
54. Shi, G., et al., *Interphase human chromosome exhibits out of equilibrium glassy dynamics.* Nat Commun, 2018. **9**(1): p. 3161.
55. Di Pierro, M., et al., *Transferable model for chromosome architecture.* Proc Natl Acad Sci U S A, 2016. **113**(43): p. 12168-12173.
56. Zhang, B. and P.G. Wolynes, *Topology, structures, and energy landscapes of human chromosomes.* Proc Natl Acad Sci U S A, 2015. **112**(19): p. 6062-7.
57. Erdel, F. and K. Rippe, *Formation of Chromatin Subcompartments by Phase Separation.* Biophys J, 2018. **114**(10): p. 2262-2270.
58. Su, X., K.E. Wellen, and J.D. Rabinowitz, *Metabolic control of methylation and acetylation.* Curr Opin Chem Biol, 2016. **30**: p. 52-60.
59. Bizhanova, A. and P.D. Kaufman, *Close to the edge: Heterochromatin at the nucleolar and nuclear peripheries.* Biochim Biophys Acta Gene Regul Mech, 2021. **1864**(1): p. 194666.
60. Briand, N. and P. Collas, *Lamina-associated domains: peripheral matters and internal affairs.* Genome Biol, 2020. **21**(1): p. 85.
61. Almassalha, L.M., et al., *Chromatin conformation, gene transcription, and nucleosome remodeling as an emergent system.* Sci Adv, 2025. **11**(2): p. eadq6652.
62. Shim, A.R., et al., *Chromatin as self-returning walks: From population to single cell and back.* Biophys Rep (N Y), 2022. **2**(1): p. 100042.
63. Carignano, M.A., et al., *Local volume concentration, packing domains, and scaling properties of chromatin.* eLife, 2024. **13**: p. RP97604.
64. Alisafaei, F., et al., *Regulation of nuclear architecture, mechanics, and nucleocytoplasmic shuttling of epigenetic factors by cell geometric constraints.* Proc Natl Acad Sci U S A, 2019. **116**(27): p. 13200-13209.
65. Damodaran, K., et al., *Compressive force induces reversible chromatin condensation and cell geometry-dependent transcriptional response.* Mol Biol Cell, 2018. **29**(25): p. 3039-3051.
66. Garate, X., et al., *The relationship between nanoscale genome organization and gene expression in mouse embryonic stem cells during pluripotency transition.* Nucleic Acids Res, 2024. **52**(14): p. 8146-8164.
67. Wang, Y., et al., *Chromatin Organization Governs Transcriptional Response and Plasticity of Cancer Stem Cells.* Advanced Science, 2025. **12**(17): p. 2407426.
68. Xu, J., et al., *Ultrastructural visualization of chromatin in cancer pathogenesis using a simple small-molecule fluorescent probe.* Sci Adv, 2022. **8**(9): p. eabm8293.
69. Perez-Gonzalez, A., K. Bevant, and C. Blanpain, *Cancer cell plasticity during tumor progression, metastasis and response to therapy.* Nat Cancer, 2023. **4**(8): p. 1063-1082.
70. Spill, F., et al., *Impact of the physical microenvironment on tumor progression and metastasis.* Curr Opin Biotechnol, 2016. **40**: p. 41-48.

71. Miroshnikova, Y.A. and S.A. Wickstrom, *Mechanical Forces in Nuclear Organization.* Cold Spring Harb Perspect Biol, 2022. **14**(1).
72. Bloom, K. and D. Kolbin, *Mechanisms of DNA Mobilization and Sequestration.* Genes (Basel), 2022. **13**(2).
73. Vidi, P.A., et al., *Closing the loops: chromatin loop dynamics after DNA damage.* Nucleus, 2025. **16**(1): p. 2438633.
74. Chatterjee, N. and G.C. Walker, *Mechanisms of DNA damage, repair, and mutagenesis.* Environ Mol Mutagen, 2017. **58**(5): p. 235-263.
75. Das, R., et al., *How enzymatic activity is involved in chromatin organization.* Elife, 2022. **11**.
76. Pommier, Y., et al., *Human topoisomerases and their roles in genome stability and organization.* Nat Rev Mol Cell Biol, 2022. **23**(6): p. 407-427.
77. Espinosa-Martinez, M., M. Alcazar-Fabra, and D. Landeira, *The molecular basis of cell memory in mammals: The epigenetic cycle.* Sci Adv, 2024. **10**(9): p. eadl3188.
78. Carter, B. and K. Zhao, *The epigenetic basis of cellular heterogeneity.* Nat Rev Genet, 2021. **22**(4): p. 235-250.
79. Dupont, S. and S.A. Wickstrom, *Mechanical regulation of chromatin and transcription.* Nat Rev Genet, 2022. **23**(10): p. 624-643.
80. Schultz, E.R., et al., *Current Advances in Genome Modeling Across Length Scales.* Wiley Interdisciplinary Reviews: Computational Molecular Science, 2025. **15**(3): p. e70024.
81. Frederick, J., et al., *Leveraging chromatin packing domains to target chemoevasion in vivo.* Proc Natl Acad Sci U S A, 2025. **122**(30): p. e2425319122.
82. Nunez, J.K., et al., *Genome-wide programmable transcriptional memory by CRISPR-based epigenome editing.* Cell, 2021. **184**(9): p. 2503-2519 e17.
83. Carnevali, D., et al., *A deep learning method that identifies cellular heterogeneity using nanoscale nuclear features.* Nat Mach Intell, 2024. **6**(9): p. 1021-1033.
84. Bhartari, A.K., V. Vinayak, and V.B. Shenoy, *Mollifier Layers: Enabling Efficient High-Order Derivatives in Inverse PDE Learning.* arXiv preprint arXiv:2505.11682, 2025.
85. Kim, H.H., et al., *O-SNAP: A comprehensive pipeline for spatial profiling of chromatin architecture.* bioRxiv, 2025.